\documentclass[preprint2,trackchanges]{aastex631}

\usepackage{graphicx}
\usepackage[suffix=]{epstopdf}
\usepackage{amsmath}
\usepackage{url}
\usepackage{xspace}
\usepackage{mathrsfs}
\usepackage{multirow}
\usepackage{tikz}

\begin{document}

\title{Searching the SN\,1987A SETI Ellipsoid with TESS}

\author[0009-0009-1019-3890]{B\'arbara Cabrales}
\affiliation{Berkeley SETI Research Center, University of California, Berkeley, CA 94720, USA}
\affiliation{SETI Institute, 339 N Bernardo Ave Suite 200, Mountain View, CA 94043}

\author[0000-0002-0637-835X]{James R.~A.\ Davenport}
\affiliation{Astronomy Department, University of Washington, Box 951580, Seattle, WA 98195, USA}

\author[0000-0001-7057-4999]{Sofia Sheikh}
\affiliation{Department of Astronomy, University of California Berkeley, Berkeley CA 94720, USA}
\affiliation{SETI Institute, 339 N Bernardo Ave Suite 200, Mountain View, CA 94043}

\author[0000-0003-4823-129X]{Steve Croft}
\affiliation{Berkeley SETI Research Center, University of California, Berkeley, CA 94720, USA}
\affiliation{Department of Astronomy, University of California Berkeley, Berkeley CA 94720, USA}

\author{Andrew P.~V.\ Siemion}
\affiliation{Berkeley SETI Research Center, University of California, Berkeley, CA 94720, USA}
\affiliation{Department of Astronomy, University of California Berkeley, Berkeley CA 94720, USA}
\affiliation{SETI Institute, 339 N Bernardo Ave Suite 200, Mountain View, CA 94043}

\author{Daniel Giles}
\affiliation{SETI Institute, 339 N Bernardo Ave Suite 200, Mountain View, CA 94043}

\author{Ann Marie Cody}
\affiliation{SETI Institute, 339 N Bernardo Ave Suite 200, Mountain View, CA 94043}

\shorttitle{The SETI Ellipsoid with TESS}
\shortauthors{Cabrales et al.}

\begin{abstract}
The SETI Ellipsoid is a strategy for technosignature candidate selection which assumes that extraterrestrial civilizations who have observed a galactic-scale event --- such as supernova 1987A --- may use it as a Schelling point to broadcast synchronized signals indicating their presence. Continuous wide-field surveys of the sky offer a powerful new opportunity to look for these signals, compensating for the uncertainty in their estimated time-of-arrival. We explore sources in the TESS continuous viewing zone, which corresponds to 5\% of all TESS data, observed during the first three years of the mission. Using improved 3D locations for stars from Gaia Early Data Release 3, we identified 32 SN\,1987A SETI Ellipsoid targets in the TESS continuous viewing zone with uncertainties better than 0.5\,ly. We examined the TESS light curves of these stars during the Ellipsoid crossing event and found no anomalous signatures. We discuss ways to expand this methodology to other surveys, more targets, and different potential signal types.
\end{abstract}

\section{Introduction}
\label{sec:introduction}

Technosignatures are any measurable property which may provide evidence of extraterrestrial technology \citep{Tarter2}. The search for extraterrestrial intelligence (SETI) is a branch of astrobiology which focuses on finding technosignatures, as their detection would provide evidence for intelligent life beyond Earth. Traditionally, targeted radio surveys have been the mainstay of SETI research \citep{Tarter}, and many SETI projects currently underway still take place in the radio band \citep[e.g.,][]{Sheikh, Ma, jlm}. However, searches in other wavelengths, such as the optical, are increasingly common \citep{Zuckerman}. New instruments in development, such as the PANOSETI project (targeting all-sky, all-the-time laser searches; \citealt{wrights}), promise to survey these underexplored territories in SETI, and expand to search strategies that are not exclusive to narrow radio frequencies. Additionally, while dedicated searches have a important place in the SETI field versus commensal and/or archival work with other astronomical projects \citep{Wright}, piggybacking from larger surveys remains relevant to the SETI field in the era of big data, e.g., the COSMIC project on the Very Large Array\footnote{\url{https://science.nrao.edu/facilities/vla/observing/cosmic-seti}}.

Modern sky surveys gather a rapidly-growing amount of information, which provides more opportunities to find technosignatures \citep{Djorgovski,Davenport2019, Lazio}. However, the development of cutting-edge instruments, such as the Vera Rubin Observatory's Large Synoptic Survey Telescope (LSST), expected to produce 15 TB of raw data each night \citep{lsst}, introduces the need for prioritization strategies which will help optimize our searches. With such a large volume of data producing a wealth of anomalies and interesting signals, we must use approaches to decide which of them are interesting in the context of SETI. 
One possibility is to use a Schelling point \citep{schelling1958strategy}, a strategy where two agents coordinate their efforts towards a common goal without communicating, using a conspicuous point of reference instead of searching randomly, which increases their odds at succeeding; these have often been applied to SETI, e.g., as magic frequencies \citep{wright2}. Another potential Schelling Point in SETI is the ``SETI Ellipsoid''. 

The SETI Ellipsoid is a geometric method of target prioritization for technosignature searches, which assumes an intentional attempt at communication from an extraterrestrial agent (ETA), as opposed to the search for signs of technology not intended for our discovery \citep{wright2}. In this model, the ETA might use a sufficiently rare and outstanding galactic-scale event, such as a nearby supernova, to synchronize a beacon indicating their presence \citep{lemarchand} hoping to catch the attention of other astronomers (i.e., us) who may be interested in studying such an event \citep{Davenport_2022}. The SETI Ellipsoid therefore helps select outliers or signals that are interesting and require further analysis or follow-up from a technosignature standpoint. This method can be used in conjunction with a variety of proposed technosignatures including brightening or dimming in a stellar light curve, narrowband radio transmissions or laser beacons. A detailed diagram and review of the SETI Ellipsoid are provided in Section 2 of \cite{Davenport_2022}.

This particular work is a proof-of-concept paper, designed to develop the idea of the SETI Ellipsoid in the context of modern instrumentation and data. In addition, we have performed a search for particular technosignatures in state-of-the-art data from the Transiting Exoplanet Survey Satellite \citep[TESS; ][]{Ricker} by looking for light curve anomalies. Many types of technosignatures could produce apparent dimming or brightening behavior in a star’s light curve, by emitting either broad or narrow wavelength signals, or blocking the star’s natural photons. 
One example could be unusual or anomalous transits in the light curve \citep[e.g.][]{zuckerman2024}, which could occur in conjunction with the Ellipsoid crossing time. Another possible signal could be an outburst that mimics the supernova profile in the light curve \citep{nilipour2023}, which would be distinct from e.g. normal stellar flares \citep{tovar}.

Our distance estimates to stars have historically been too imprecise for a geometric search of this kind, and have presented the greatest obstacle to the SETI Ellipsoid technique \citep{makovetskii1977}. However, the Gaia mission produced improved distance uncertainties by calculating parallaxes with a precision $\sim 100 \times$ better than its predecessor Hipparcos \citep{gaia1}. It is now possible to apply the SETI Ellipsoid technique as a method of target prioritization. For example, there were 134 SETI Ellipsoid candidates from the Gaia Catalog of Nearby Stars (GCNS) alone within 0.1 ly of the SN\,1987A Ellipsoid in 2022 \citep{Davenport_2022}.

Gaia observations are sporadic and sparse over years \citep{Hodgkin}, which makes their light curves suboptimal for short timescale technosignature monitoring \citep{nilipour2023}. Additionally, distance uncertainties propagate into uncertainties in the expected arrival time for a potential technosignature. Optimal windows of observation are proportional to the distance uncertainties of stars --- which work out to about a couple of months for nearby targets. Thus the Gaia mission can be used to identify interesting targets and times for observation, but on its own, it can only provide sparse light curves with which to search for the hypothesized synchronous technosignature. Here, we work around this shortcoming by combining Gaia distance estimates from Early Data Release 3 (EDR3) and continuous observations from TESS, to search for anomalies in light curves which span longer periods of time.

Unlike its forerunner, Kepler, the TESS survey covers almost the whole sky, surveying one near-full hemisphere a year by shifting its field of view to 13 different sectors in that time. Furthermore, these 13 fields overlap at the ecliptic poles, providing yearlong coverage to targets located in this area, referred to as the continuous viewing zone (CVZ). Searching for CVZ targets on the SETI Ellipsoid will ensure access to one year's worth of uninterrupted light curve data for each source. This will provide enough information to establish a target's baseline behavior for anomaly detection strategies, and enough leeway to account for the uncertainty of a hypothetical signal's time-of-arrival.

In this work, we show the advantages of using a continuous wide-field survey, such as TESS, to monitor targets on the SETI Ellipsoid. In Section \ref{sec:tess_and_gaia}, we describe our procedure for retrieving the 3D coordinates of the CVZ sources. In section \ref{sec:select}, we discuss the parameters we used to filter our initial data sample and describe the Ellipsoid algorithm for SN\,1987A, allowing us to select the final targets which we analyze in Section \ref{sec:targets}, focusing especially on the targets' TESS light curves and their features. Finally, we summarize our findings and consider future work in \ref{sec:discuss}.

\section{Data Selection: TESS and Gaia}
\label{sec:tess_and_gaia}
Jointly, TESS and Gaia are well-suited surveys to carry out the SETI Ellipsoid technique due to the complementary features of the missions; TESS offers densely sampled light curves over long time periods, and Gaia provides the most precise distances to stars ever computed. Additionally, their combined focus on nearby sources --- which have smaller distance uncertainties --- reduces the average time range that we need to monitor for, since the error on the arrival time of a synchronized signal from a star is directly impacted by the star's distance uncertainty. Given that the maximum timescale for TESS's continuous photometric coverage is one year, nearby stars with timing uncertainties that fit within this timescale are particularly good targets.

\begin{figure*}[!ht]
\centering
\includegraphics[width=7 in]{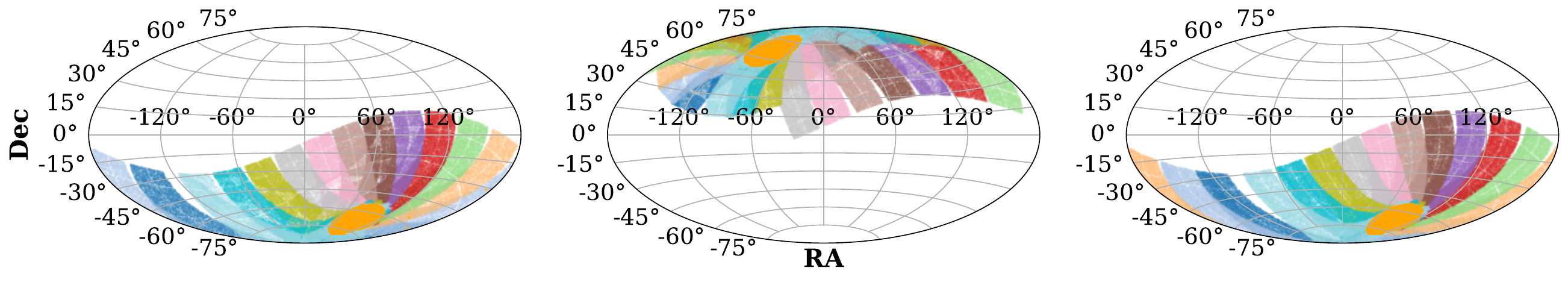}
\caption{TESS data obtained during Cycle 1 (left), Cycle 2 (center), and Cycle 3 (right) on the celestial plane. Common RA and Dec labels are displayed for the entire Figure. Sources are color-coded by sector, highlighting the geometry of the TESS field of view and the overlap of all 13 sectors --- per cycle --- at the ecliptic poles (the CVZ). The circular area (bright orange) near the pole on each plot is composed of the targets observed in at least 11 overlapping sectors during that cycle, and shows the data sample from which SETI Ellipsoid targets were selected.}
\label{fig:tess-cvz}
\end{figure*}

\subsection{TESS and the CVZ}
\label{ssec:tess_cvz}
The TESS mission succeeded Kepler in the search for exoplanets, focusing on stars 100$\times$ brighter and 10$\times$ closer to Earth --- on average --- than the earlier mission. As a result, TESS's coverage is mostly constrained to stars that lie within 200--300 lyr of Earth as opposed to Kepler's average of $\sim$3000 lyr \citep{Jenkins}. Moreover, while the Kepler prime mission field of view surveyed a fixed 100 deg$^2$ area near the constellation Cygnus, the TESS field of view shifts every 27.4 days, allowing it to complete an all-sky survey in 2 years. \citep{Koch, Ricker}. 

One 27-day continuous coverage of TESS's 24$^{\circ}$ $\times$96$^{\circ}$ field of view is known as an observing sector. One year of coverage --- hereafter referred to as a Cycle --- contains 13 sectors and observes most of one hemisphere \citep{Ricker}. In this work, we use the 2-minute cadence data from the first three cycles, completed during 2018 through 2021. As shown in Figure \ref{fig:tess-cvz}, Cycles 1 and 3 --- which ran during 2018--2019 and 2020--2021 respectively --- provided coverage of the southern hemisphere sky, whereas Cycle 2 --- active during 2019--2020 --- surveyed the northern hemisphere. 

During these three cycles, 329,176 unique stars were observed, but most of these sources were only present in one sector per cycle. This is apparent in Figure \ref{fig:tess-cvz}. We are mainly interested in targets that appear in multiple sectors, and therefore have long spans of continuous photometric coverage, since this helps us account for the uncertainties in the expected arrival time of a presumed signal, and provides multiple month-long, high-cadence light curves to use for anomaly detection. Consequently, we focused our search on stars which were present in at least 11 sectors in a given cycle, such that there are between 300 and 356 days worth of data for each target. This criterion shrinks our sample by 96\%, resulting in 14,614 stars to select SETI Ellipsoid targets from, and guarantees that we search the entire CVZ, as well as nearby fields with very good coverage. The distribution of the reduced data sample on the celestial plane is shown in bright orange in Figure \ref{fig:tess-cvz}; we will refer to this sample simply as the ``CVZ'' in the following text. 

Since Cycles 1 and 3 surveyed the same hemisphere, 48,744 southern-sky sources were observed during both time periods, two years apart; 2,493 of which were in the CVZ. Note that, although roughly the same area was surveyed, only about 15\% of targets overlap. This is because we are using the TESS 2-minute targets. These targets were selected through proposals, which saw a 500\% increase in the first extension of the mission (Cycles 3 and 4). \footnote{\url{https://heasarc.gsfc.nasa.gov/docs/tess/extended.html}} The Full Frame Image (FFI) data from TESS include far more targets than those considered in this work. We further discuss FFIs in Section \ref{sec:discuss}.  As will be described in Section \ref{sec:select}, timing is essential for this technique; thus, a source that might be trivial in 2019 could be a target of interest in 2021. Accordingly, we decided to independently consider the 2,493 overlapping targets in both cycles, increasing our database of CVZ targets to 17,107 rows.

\subsection{Distances from Gaia EDR3}
\label{ssec:gaia_dist}
The SETI Ellipsoid technique of target selection can only be applied when the distances to the surveyed stars are precise and accurate. Gaia's EDR3 provides a significant improvement in parallax measurement precision over the previous Data Release 2 (DR2), leading to median parallax uncertainties of 0.02--0.03\,mas for bright stars (G $\le$ 14\,mag), a 40\% decrease from Gaia DR2, which had median uncertainties of 0.04\,mas for stars of the same magnitude \citep{edr3, Lindegren2,Lindegren}. With this unprecedented parallax precision, \cite{Bailer-Jones} estimated geometric and photogeometric distances for 1.5 billion stars. As shown by \cite{Davenport_2022}, the catalog of Gaia EDR3 geometric-probabilistic distances from \cite{Bailer-Jones} can be effectively used to make precise estimates for the expected arrival time of a presupposed synchronized signal using the SETI Ellipsoid method for a well-documented source event. Here, we use the same procedure of target selection but for TESS objects, focusing on the 14,614 CVZ targets chosen in \ref{ssec:tess_cvz}. We did not limit these targets by any particular parameter (e.g., stellar type, binarity, or habitability). A system could be used for the SETI Ellipsoid technique regardless of whether it is habitable --- in fact, it might be preferable for the ETA to place the signalling technology in a uninhabited system.

\subsection{Crossmatching the CVZ targets with EDR3 distances}
\label{ssec:xmatch}

To assign each target its corresponding Gaia ERD3 probabilistic distances, we used the cross-match service provided by CDS, Strasbourg \citep{xmatch}.

We merged our TESS data with the Gaia EDR3 distances database using the targets' celestial coordinates within a radius of $1\arcsec$. During the crossmatching process of the CVZ sample, 725 sources were dropped, likely due to a disagreement greater than the $1\arcsec$ threshold between the TESS and Gaia RA and Dec values. We did not attempt to recover these missing targets, which correspond to 4\% of the initial CVZ list. Thus, the final data sample contained 16,382 rows of TESS CVZ sources with Gaia EDR3 distance estimates.

\section{Selecting Targets on the SN1987A SETI Ellipsoid }
\label{sec:select}

The addition of Gaia distances to our sample enables us to compute the distance to a SETI Ellipsoid for any given target at any given time. This allows for the selection of targets which are optimally placed for a communication attempt coordinated with astronomical phenomena according to the SETI Ellipsoid framework.

\subsection{The Mathematics of the SETI Ellipsoid}
\label{ssec:ellipsoid_review}

As noted in Section \ref{sec:introduction}, the SETI Ellipsoid is a method of target selection that works under the assumption that an extraterrestrial civilization or agent would actively try to draw attention to themselves by broadcasting a signal soon after first witnessing a notable astronomical event \citep{makovetskii1977, lemarchand}.

The speed of light, $c$, constrains our targets of interest to a particular region of space at a particular point in time. We consider stars located at the precise distances from Earth that allow for an electromagnetic signal --- sent in coordination with the astronomical event --- to reach Earth at any given time, with the intent of aligning our monitoring efforts accordingly. 

The region of interest is the ellipsoid whose foci are at the source event and at Earth. The ellipsoid grows with time as new stars observe the event, and is defined by:
\begin{equation}
    d_1 + d_2 = 2\, \mathscr{A} = 2\, \mathscr{C} + T
\label{eqn}
\end{equation}
\noindent
where ($d_1$) is the distance from the Earth to a target laying on the ellipsoid, ($d_2$) is the distance from the event to the target, $\mathscr{A}$ is the semi-major axis of the ellipsoid, $\mathscr{C}$ is the distance from each foci to the center of the ellipsoid, and $T = c\, t$ is the time since Earth observed the event times the speed of light. Therefore given a synchronizing event with a known distance, and a sample of field stars with distances, we can compute their Ellipsoid crossing times as a function of time using Equation \ref{eqn}.

\subsection{SN\,1987A as the synchronizing event}
\label{ssec:sn1987a}

Ideal events for the Ellipsoid technique are fixed in space and time, relatively uncommon in our galactic neighborhood, and bright enough to be used as a synchronizing beacon on a galactic scale \citep{lemarchand}.

\begin{figure}[!ht]
\centering
\includegraphics[width=3.3in]{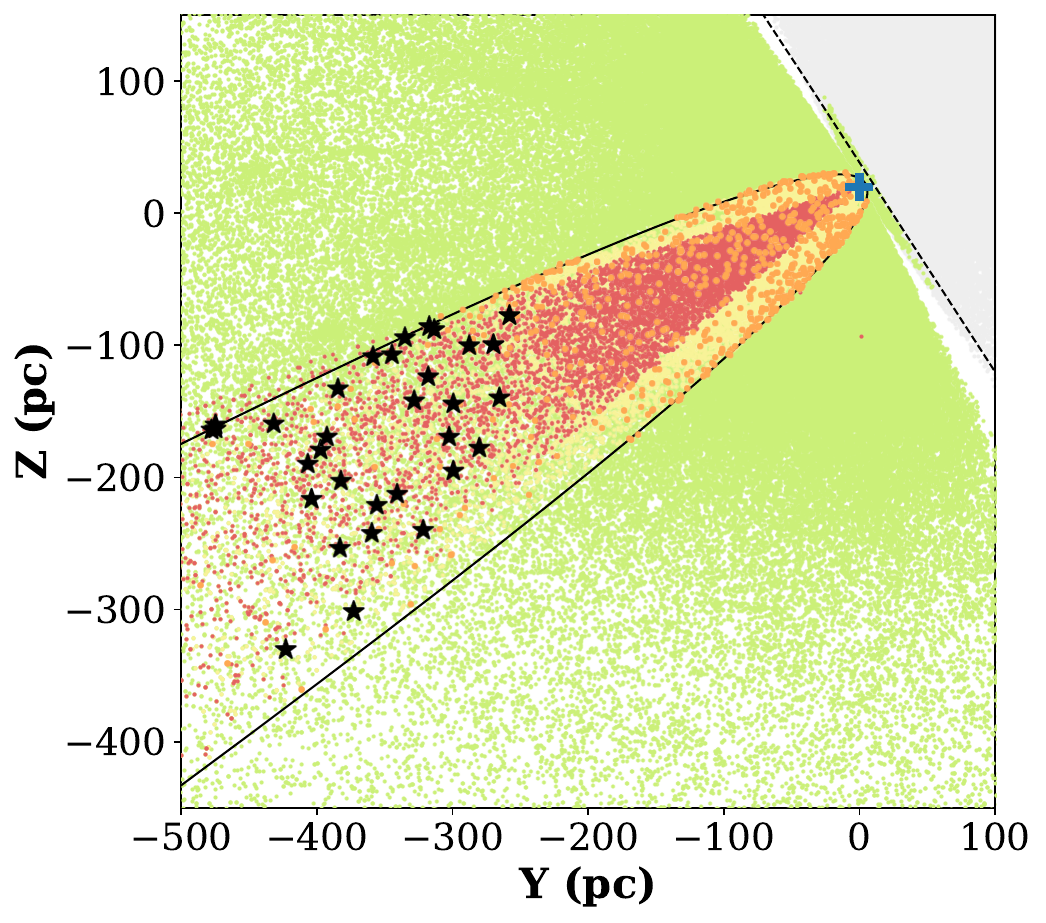}

\caption{
Galactocentric Y-Z plane for TESS stars, color coded to show the location of the southern CVZ, and their designation within the SETI Ellipsoid framework for SN\,1987A. 214,995 stars have seen SN\,1987A (dashed line) but are outside of the ellipsoid (green points). 97,514 stars have not observed SN\,1987A (gray points). 23,148 stars are inside the SN\,1987A Ellipsoid (yellow points), implying that any coordinated beacon has already passed the Earth. 736 stars are laying on the SN\,1987A ellipsoid (orange points). The 7,850 southern CVZ stars are shown in pink. 37 targets are both on the ellipsoid and in the southern CVZ (black stars) and are our targets of interest. The contour of the SETI Ellipsoid in the YZ plane is shown as a solid line. The sun is also plotted (blue cross).}
\label{fig:ellipsoid}
\end{figure}

Although other events are worthy of examination as well \citep[e.g., galactic novae, as explored by][]{Davenport_2022}, supernova outbursts are particularly good source events to place at one of the foci within the Ellipsoid framework, and SN\,1987A is an ideal Schelling point for our purposes. First, its distance to Earth is very well-documented with a low uncertainty. We used the distance estimate computed by \citet{panagia} of 51.4 $\pm$ 1.2 kpc, where the 1.2 kpc uncertainty corresponds to a 2.3\% error. Secondly, SN\,1987A is the closest supernova to Earth since SN\,1604. Given that it is relatively nearby in space-time, SN\,1987A has a thin associated Ellipsoid which closely surrounds Earth, ensuring that many targets of interest will lie within relatively short distances and therefore have low distance uncertainties, resulting in low timing uncertainties for the arrival time of a synchronized signal. Lastly, the location of SN\,1987A in the Large Magellanic Cloud (LMC) is optimal for this work due to the cloud's proximity to the south ecliptic pole, which is the area surveyed by TESS's CVZ in Cycles 1 and 3. The polar location, along with the current thin shape of the ellipsoid, allows for the intersection of a significant portion of SN\,1987A's ellipsoid with the southern TESS CVZ cone, making it highly likely to find several targets of interest within our previously selected data sample for two of the three cycles considered. In contrast, we expect little to no Ellipsoid candidates on the northern hemisphere. This can be appreciated in Figure \ref{fig:ellipsoid}.

\subsection{The SN\,1987A Ellipsoid}
\label{ssec:sn187aellipsoid}
The list of TESS CVZ sources found in Section \ref{ssec:tess_cvz} contains 14,614 individual objects and 17,107 rows of data due to sources which were observed in two different cycles. After the cross-match with Gaia distances performed in \ref{ssec:xmatch}, we know both where these targets are located in space, and when they were observed by TESS. Solving equation \ref{eqn} for a desired time allows us to calculate the distance of our CVZ targets to the Ellipsoid at that time.

The algorithm we developed uses the start and end dates of observation for each target's first and last sectors respectively, obtained from NASA's TESS observation times site\footnote{\url{https://heasarc.gsfc.nasa.gov/docs/tess/sector.html}}. The mid-time of observation is then calculated, and since most of the targets in our sample were observed in all TESS sectors, this mid-time is the same for every 13-sector target in each cycle as shown in Table \ref{tab:midtimes}.

\begin{table}[h!]
\begin{tabular}{|c|l|}
\hline
Cycle & Mid-Time Date \\
\hline
1 & 2019 January 20\\
2 & 2020 January 10\\
3 & 2020 December 28\\
\hline
\end{tabular}
\caption{Mid-times for 13-sector TESS observations taken during the three cycles considered in this work.}
\label{tab:midtimes}
\end{table}

We then use the right-hand side of equation \ref{eqn} to find the length of the semi-major axis of the ellipsoid, $\mathscr{A}$, at these times. According to equation \ref{eqn}, a target is on the ellipsoid if 
\begin{equation}
    d_1 + d_2 - 2\, \mathscr{A} = 0
\label{eqn2}
\end{equation}
\noindent

however, we have decided to select targets which are within reasonable distance to the ellipsoid as opposed to exactly on it, given the uncertainties involved in the process of finding a target strictly on an infinitely thin surface. Other uncertainties might also play a role, such as the error in the distances to the sources, explored in Section \ref{ssec: uncertainties}, or the potential delay in the broadcast of a burst from an ETA, discussed in Section \ref{sec:discuss}. We allowed for some leeway in the ellipsoid distance by creating a tolerance parameter, \texttt{etol}, a positive value in ly which determines the distance-to-the-ellipsoid threshold for target selection. This adds thickness to the ellipsoid, essentially turning it into an ellipsoidal shell defined by:

\begin{equation}
    |d_1 + d_2 - 2\, \mathscr{A}| < \texttt{etol}
\label{eqn3}
\end{equation}
\noindent

The same steps were followed by \cite{Davenport_2022}, where the selected tolerance was \texttt{etol} = 0.1\,ly, to make use of Gaia's highest timing precisions. However, as \citeauthor{Davenport_2022} state, the value of this threshold can and should be picked based on the specific monitoring campaign's properties. For this work, we have selected \texttt{etol} = 0.5\,ly to match the approximate duration of one TESS cycle (1 year). Choosing targets 0.5\,ly away from the mid-cycle ellipsoid helps account for targets which would have been on it at the beginning of said cycle --- 0.5 years before mid-time --- or at the end --- 0.5 years after mid-time. This is not a perfectly precise threshold since the ellipsoid grows slightly slower than the speed of light near the foci, resulting in a decreasing eccentricity over time such that the ellipsoid becomes a spheroid at $t = \infty$. This widening effect is illustrated in Figure 5 of \cite{Davenport_2022}. Due to this ellipsoid rate of growth differential, a 0.5\,ly tolerance potentially includes targets near Earth or near SN\,1987A that did not intersect the Ellipsoid exactly during the TESS's observing cycle. Note that this widening effect occurs over centuries, and is likely negligible over timescales of individual missions.

After running our algorithm for the 17,107-row list of CVZ sources, we obtained 12 targets of interest laying within 0.5\,ly of the SN\,1987A ellipsoid as of 2019 January 20 --- mid-time of Cycle 1 --- and 24 targets of interest as of 2020 December 28 --- mid-time of Cycle 3, resulting in 36 selected TESS objects. As expected per the discussion in Section \ref{ssec:sn1987a} no targets of interest were found in the northern hemisphere, surveyed during Cycle 2. Accordingly, we did not extend our search to TESS's Cycle 4, which also observed this region of the sky.

\subsection{Incorporating Positional Uncertainties}
\label{ssec: uncertainties}

The selection of the 36 targets of interest from our list of 17,107 CVZ sources described in Section \ref{ssec:sn187aellipsoid} was computed considering only their 50\% probability geometric distances, known as $r_{med}$, obtained from \cite{Bailer-Jones}, as well as the reported distance to SN\,1987A from \cite{panagia}, and the celestial coordinates from the TESS two-minute data, not accounting for errors for any of these quantities. However, these errors directly impact the uncertainties in the intersection time between the Ellipsoid and any target of interest, which consequently, establishes a window for the expected time-of-arrival of a synchronized signal.

\begin{figure}[!ht]
\centering
\includegraphics[width=3.3 in]{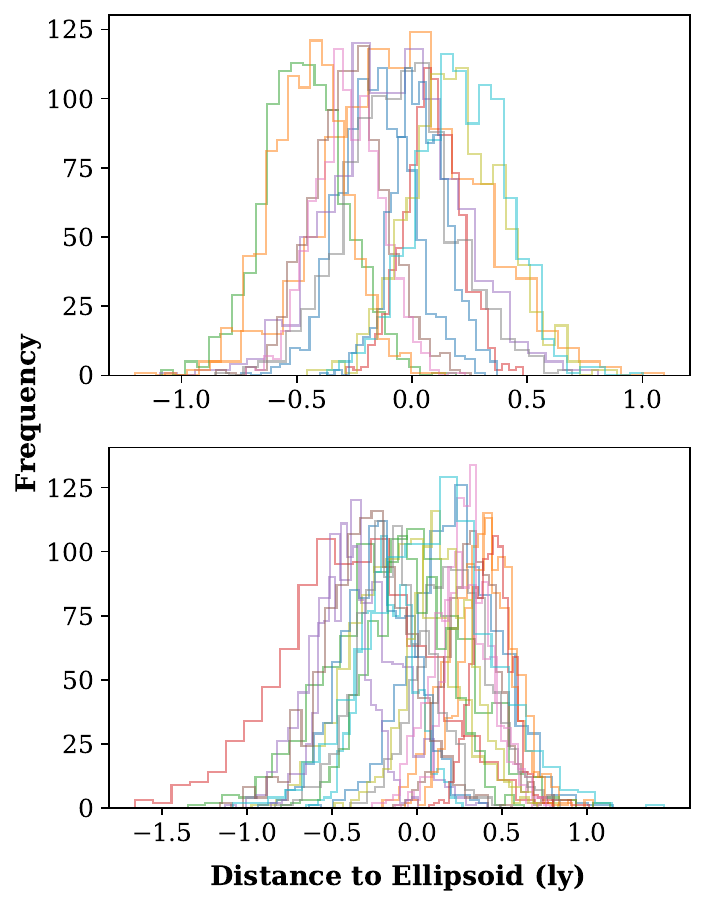}
\caption{Histogram of simulated distances of the Cycle 1 Ellipsoid candidates (top) and the Cycle 3 Ellipsoid candidates (bottom) to the SN\,1987A ellipsoid using a Monte Carlo method of random draws in a Gaussian distribution calculated from Gaia’s distance errors. Common Distance to the Ellipsoid and Frequency labels are displayed for the entire Figure. Even when including all sources of uncertainty, all 32 of these targets have a standard deviation smaller than 0.5\,ly. The peaks of these distributions are also within 0.5\,ly, as expected from the \texttt{etol} value chosen in Section \ref{ssec:sn187aellipsoid}.}
\label{fig:mc}
\end{figure}

We investigated how much the error of each of these quantities affects the distance of our targets to the ellipsoid to a) assess which one of these uncertainties will have the greatest impact on our calculations, b) to eliminate targets with very large uncertainties, i.e. uncertainties larger than TESS's yearlong coverage and c) to find the previously mentioned ``window of interest'' to monitor, which corresponds to the range of time where we expect to see an anomaly in the light curves of our targets of interest coordinated with their first detection of SN\,1987A.
To accomplish this, we incorporated the RA, Dec, SN\,1987A distance and target distance uncertainties into our original selection algorithm by using a Monte Carlo simulation.
We recalculated the distance to the ellipsoid for each selected target by drawing 1000 random Gaussian-distributed samples using the standard deviations for the parameters mentioned above. For the target distances, we used Bailer-Jones' 16\% and 84\% distance probabilities , $r_{lo}$ and $r_{hi}$ respectively, such that $\sigma_r = (r_{hi} - r_{lo})/2$. For the SN\,1987A distance, we used the uncertainty from \cite{panagia}, $\sigma_{SN}$ = 1.2kpc. For the celestial coordinates, RA and Dec, we used the errors from \cite{Bailer-Jones}, and \texttt{Astropy} \citep{astropy:2022} to assign celestial coordinates to each draw using the angle standard deviation\footnote{We note $\sigma_{RA}$ has the typical spherical coordinate dependence on $\cos(Dec)$.} we defined as $\sigma_{\angle}=(\sigma_{RA}^2+\sigma_{Dec}^2)^{1/2}$.
We first ran the Ellipsoid calculation including the uncertainty for each parameter in isolation in order to determine which error has a greater effect on the final results. We found that the average error that propagates from the uncertainty of the celestial coordinates to the distance-to-the-ellipsoid uncertainty corresponds to 0.073\,ly, while the uncertainty in the distance to SN\,1987A only contributed 0.006\,ly to the mean error. As expected, the dominant contribution to the distance-to-the-ellipsoid uncertainties propagates from the \cite{Bailer-Jones} error of the distance to the targets, which works out to 0.193\,ly in average. The final simulation did, however, consider the error in all the evaluated parameters. 

A histogram illustrating the results of the Monte Carlo simulation for our ellipsoid targets per cycle is shown in Figure \ref{fig:mc}. We used the standard deviation of these results to assign a distance-to-the-Ellipsoid uncertainty to each target in light years, which is equivalent to finding their ellipsoid timing uncertainty, $\sigma_{et}$,  in years. This uncertainty will indicate the length of the window of interest, defined earlier as the best time to look for abnormalities in the light curves of a selected target.   

All targets in Cycle 1 had timing uncertainties below 0.3\,yr due to their proximity to Earth. However, some targets in Cycle 3 happened to lay farther along the Ellipsoid towards SN\,1987A. The most distant targets in Cycle 3 had uncertainties of up to 13\,yr, and with TESS's continuous coverage lasting about one year, we decided to set an ellipsoid timing uncertainty threshold and only consider targets below $\sigma_{et}$ = 0.5\,yr, which matches the tolerance parameter we established earlier and guarantees that most of the window of interest for any given target will occur during TESS's coverage, even when considering a distance-to-the-ellipsoid value close to 0.5\,ly. Only 4 targets were dropped using this criterion, reducing the number of ellipsoid targets to 32. The histogram in Figure \ref{fig:mc} shows the distribution of distances to the ellipsoid for the final 32 targets. In Table \ref{bigtbl} we list all 32 of our final TESS targets, their coordinates, and the estimated SN\,1987A Ellipsoid crossing times.

\section{SN\,1987A SETI Ellipsoid Targets}
\label{sec:targets}

\begin{figure}[!t]
\centering
\includegraphics[width=3.3in]{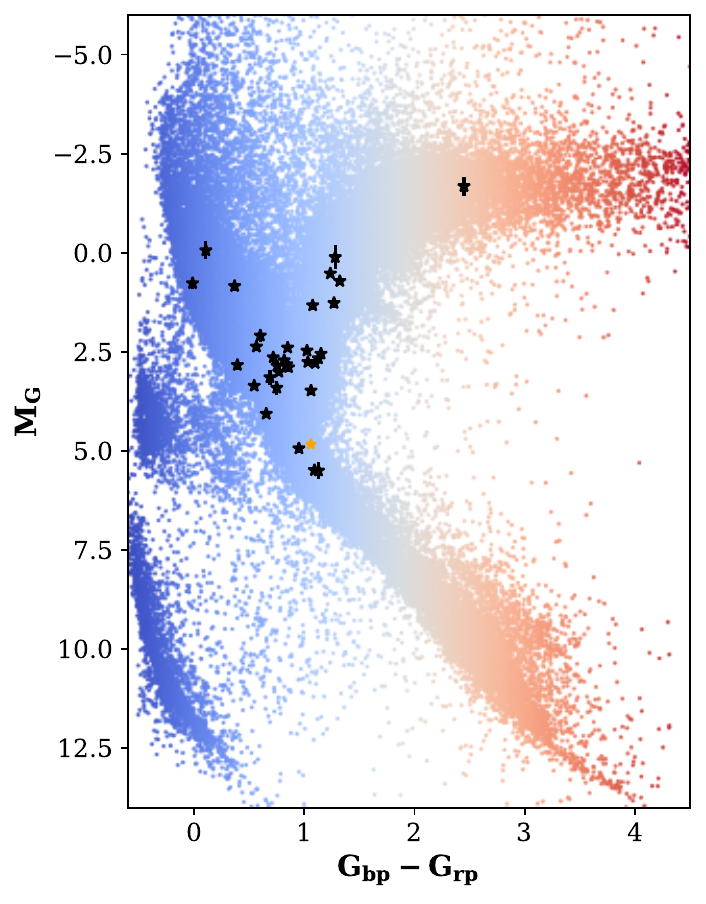}
\caption{
Gaia ($G_{bp}-G_{rp}$, $M_G$) color-magnitude diagram (CMD) for the 32 stars that we find intersect the SN\,1987A Ellipsoid (black stars with error bars). The entire sample of 329,176 TESS 2-minute targets is shown for reference, with point colors indicating the Gaia colors. The theoretical location of the Sun is also shown (orange star). Ellipsoid targets in TESS Cycles 1--3 include main sequence, sub-giant, and giant branch stars. Three stars are found with near-Solar locations in the CMD.
}
\label{fig:cmd}
\end{figure}

In Figure \ref{fig:cmd} we present the Gaia color--magnitude diagram for the 32 star systems we have identified as being within 0.5\,ly of the SN\,1987A Ellipsoid during their TESS 2-minute cadence observations. For context, we also show the entire sample of the 329,176 TESS sources surveyed at 2-minute cadence with cross matches to Gaia DR3 in the first three cycles of the mission. As expected, stars at a wide range of stellar evolutionary stage are found to intersect the Ellipsoid, with multiple turn-off and giant branch stars included, along with 3 solar-type main sequence stars, and several apparent high-mass stars. The SETI Ellipsoid framework provides a useful method for identifying all types of nearby stars for SETI monitoring from surveys like TESS, which is important for exploring all possible types of technosignatures \citep{lacki2020}.

\vspace{0.5in}

\subsection{Finding the Ellipsoid Crossing Times}
\label{ssec: xt}

While using Equation \ref{eqn3} with \texttt{etol} = 0.5\,ly is a computationally efficient method of target selection and finding the value of $\sigma_{et}$  is equivalent to calculating the length of the window of interest, we still need to find exactly when this window occurs for each target. To do this, we set out to compute the ``Ellipsoid crossing times'' for each selected source, that is, the time when the Ellipsoid and a given target are expected to meet.

\begin{figure}[!t]
\centering
\includegraphics[width=3.3 in]{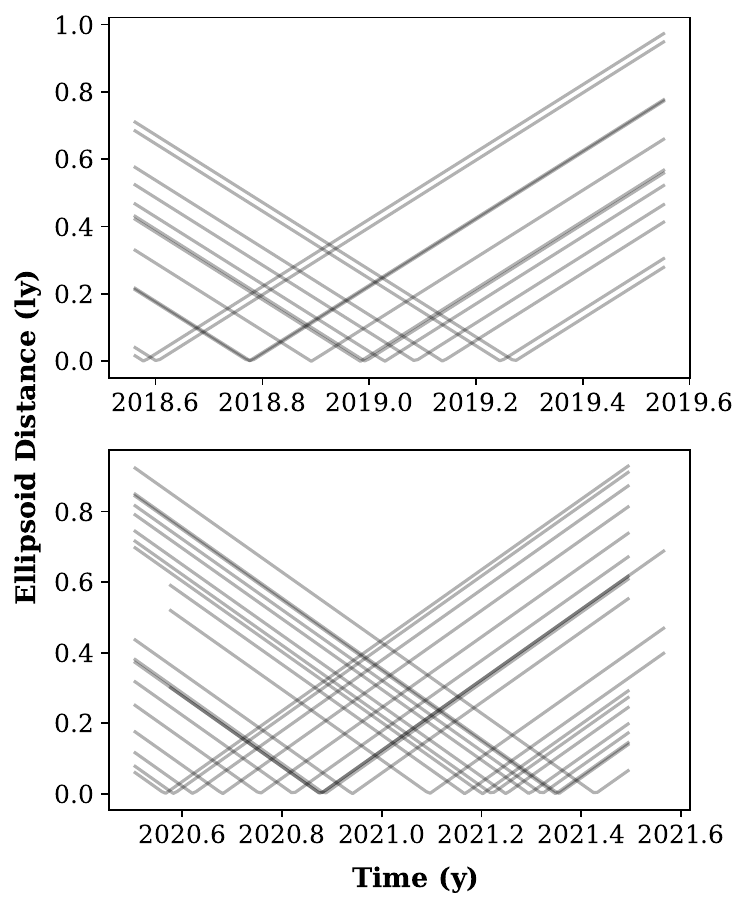}
\caption{Distance to the ellipsoid in time for targets on Cycle 1 (top) and 3 (bottom). The slope of these lines corresponds to 99.9\% the speed of light. The Ellipsoid crossing times are the zeroes in each line. Targets with lines that start and end later than the rest belong to targets whose first observation was not in Cycle 3's first sector.}
\label{fig:xtimes}
\end{figure}

To do this, we used the following rearrangement of Equation \ref{eqn2},

\begin{equation}
    d_1 + d_2 - 2\, \mathscr{A} = E_{dist}
\label{eqn4}
\end{equation}
\noindent

where the left hand side is now equal to $E_{dist}$, the distance of the target to the Ellipsoid at a given time in pc. We used Gaia's $r_{med}$ distance as our $d_1$, and the date of each candidate's first TESS observation to get $\mathscr{A}$. We then calculated the distance to the ellipsoid 130 subsequent times in steps of 2.8 days for a total of 364 days, the approximate duration of one TESS cycle. This process is illustrated in Figure \ref{fig:xtimes}, where we find the Ellipsoid crossing time for each of the 32 targets of interest. As previously mentioned, the uncertainty of this value is equal to $\sigma_{et}$, calculated in Section \ref{sec:select}.

Additionally, in Section \ref{ssec:sn187aellipsoid} we found that depending on the location of the targets in space, while their crossing times might fall within the \texttt{etol} = 0.5\,ly tolerance, the actual Ellipsoid crossing times may not be within the TESS observing Cycle. As we can see in Figure \ref{fig:xtimes}, this was not an issue and all the targets did intersect the ellipsoid within their respective TESS cycles. Furthermore, the slopes of the lines (i.e. the Ellipsoid distance over time) are $\sim 0.999 c$, showing that, as expected, any apparent super-luminal projection effects are negligible at the scale of $\sim 1$ year.

\vspace{0.5in}
\subsection{Light curves and search for anomalies}
\label{ssec: light curves}

Using the \texttt{lightkurve} package \citep{lightkurve}, we obtained all the 2-minute cadence PDCSAP (Pre-search Data Conditioning Simple Aperture Photometry) TESS light curves for each candidate. PDCSAP fluxes are produced in the TESS science pipeline by identifying and removing long-term systematic trends from the SAP (Simple Aperture Photometry) flux, which is generated by adding all pixel values in a predetermined aperture \citep{Morris}. After retrieving the PDCSAP data, we normalized the flux in each sector, subsequently stitching all the light curves together for each candidate. This allowed us to properly characterize the year-long photometric behavior of each target of interest to establish a benchmark from which to search for deviations or anomalies. We then performed additional flux-outlier removal using the built-in sigma-clipping function in the \texttt{lightkurve} package. 

The previously-computed Ellipsoid crossing times allow us to identify the range of time where we would expect to see a technosignature appear if an extraterrestrial civilization had sent a signal immediately after observing SN\,1987A. Consequently, we looked for anomalies in all the retrieved light curves, prioritizing the window of interest given by the uncertainties in the crossing times discussed in Section \ref{ssec: uncertainties}.

\begin{figure*}[ht!]
\centering
\includegraphics[width=7 in]{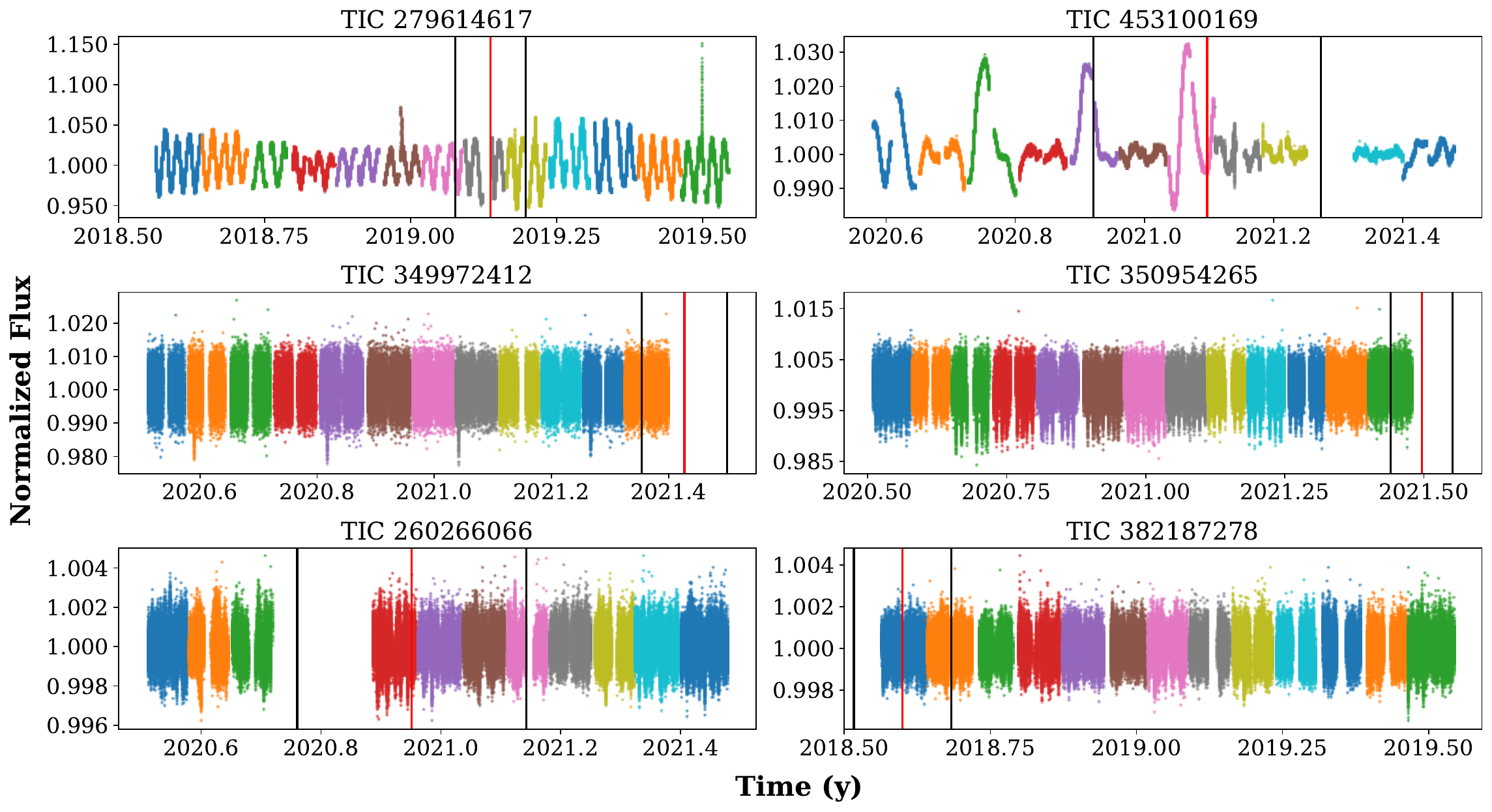}

\caption{Normalized light curves for six selected SN\,1987A Ellipsoid crossing targets that exhibit interesting behavior related to stellar variability, described in the text. Common Time and Flux labels are displayed for the entire Figure. Each light curve is color-coded by TESS sector to aid in viewing. The vertical red line in each plot indicates the crossing time with the SN\,1987A Ellipsoid. The vertical black lines show the uncertainty window of the crossing time calculated in Section \ref{sec:select}. None of these objects show any change in their light curve behavior coincident with the Ellipsoid crossing times.}
\label{fig:light curves}
\end{figure*}

We first visually inspected the light curves of all 32 candidates, from which we report no unexplained anomalies. As in \citet{nilipour2023}, we looked for changes in the stellar variability coincident with the Ellipsoid crossing times as a possible sign of anomalous behavior. However, no changes in the underlying stellar variability properties were apparent in the sample.

Changes that would have warranted further inspection include unusual brightening or dimming signals in the wide TESS band light curves, which may stand out due to amplitude or morphology. Assuming intentionality, we expect magnitude changes of at least 2--10\% in order to stand out from noise. This would indicate either the presence of megastructures of sizes comparable to several Jupiter radii, or very high energy emissions, which are possible for lasers \citep{kipping}. However, it is key to consider these amplitude changes in the context of the usual light curve behavior of the source; a brightening event in a star that has flares is less likely to stand out unless it exceeds the typical magnitude change due to stellar activity by a significant amount or has an anomalous shape. This example is particularly relevant, as a flare star is further discussed later in this section.

As for light curve changes in shape, a signal that mimics the light curve of the supernova is one possible morphology to look for, as noted by \cite{nilipour2023}. Unfortunately, the long timescale PDCSAP calibration of TESS 2-minute data is not currently reliable for recovering low amplitude slow signals, though work is underway with the TESS FFI data \citep{unpopular}. Therefore, we focus on shorter-term flares and dips in this work.

As expected from the range of stellar evolutionary stage of our sample shown in Figure \ref{fig:cmd}, the TESS light curves exhibit a variety of variability amplitudes and timescales. This includes rotational modulation, pulsation, flares, and eclipses. In Figure \ref{fig:light curves} we present a sample of six of our Ellipsoid crossing candidates from Table \ref{bigtbl} that demonstrate a range of stellar variability behaviors. As noted earlier, all of these targets show no discernible change in their normal variability patterns coincident with the Ellipsoid-crossing times. 
TIC\,279614617, also known as CPD-57 1131, is a RS CVn Variable with high amplitude (9.5\%) modulations from the 7.4 day rotational period \citep{martinez}. This system also exhibits several large flares throughout its TESS light curve, one of which occurs during the Ellipsoid window of interest. However, these instances are consistent with both the magnitude changes and the ``fast rise, slow fall'' behavior of known stellar flares, which are typical for these magnetically active binary systems \citep{tovar}.  
TIC\,453100169 (HD\,63669) is a MIII giant star with long period quasi-periodic modulations. The shapes of these modulations are visually compelling, but stay consistent throughout the yearlong TESS light curve with no detectable changes during the window of interest.  
TIC\,349972412 and TIC\,350954265 are eclipsing binary systems. We calculated the orbital periods of these systems using the Box Least Squares (BLS) periodogram \citep{BLS}, from which we report a 84.5 day period and 3.3 day period, respectively. A detailed characterization of these binary star systems --- which show no contextual anomalies --- is beyond the scope of this work.
TIC\,260266066 (HD\,43902) and TIC\,382187278 are nearby giant stars (d$\sim$621 pc and 432 pc, respectively), with apparent asteroseismic oscillations in their light curves. No unusual behavior, coordinated with SN\,1987A or otherwise, is detected.

We also used an outlier detection algorithm, conceptually based on DBSCAN \citep{ester} and built with \texttt{Scikit-Learn} \citep{scikit-learn}, that was developed specifically for detecting anomalous stellar light curves \citep{Giles1, Giles2} and applied to TESS light curves derived from the full frame images in Cycles 1 and 2 (Giles et al. {\it in preparation}). This generated a ``weirdness score'' to each target in every sector of data. This analysis was motivated by a search for unusual dimming patterns that could be indicative of the transit of an artificial structure. While inspired by systems like Boyajian's star, this outlier detection method is signal-agnostic, specifically with the purpose of avoiding assumptions related to what a technosignature might look like. Therefore, both intentional and unintentional technosignatures may be flagged as ``anomalous'' using this classifier. 

The outlier detection algorithm was run on the 12 Ellipsoid crossing targets from Table \ref{bigtbl} with crossing times in Cycle 1. None of these targets stood out as candidates needing further investigation related to anomalous light curve behaviors associated with possible technosignatures. The algorithm flagged TIC\,279614617 as a target with strong sinusoidal flux over time, ranking it within the top 2.3\% most anomalous TESS objects for sectors 2-5 and 7-13. TIC\,382187278 also stood out in Sectors 4 and 8 as being in the top 4.2\% most anomalous targets in these Sectors. No changes in the stellar variability occurred across these data.

\section{Discussion and Conclusions}
\label{sec:discuss}

We have demonstrated a practical application of the SETI Ellipsoid technique to examine archival data in an attempt to identify light curve anomalies that might be indicative of technosignatures synchronized with astronomical events. Building from Gaia's EDR3 highly precise distance estimates, we have showcased the feasibility of cross matching these data to other time domain surveys, such as TESS, to expand on the monitoring and anomaly detection aspect of this kind of SETI search.

TESS's continuous viewing zone provides yearlong light curves for targets near the ecliptic poles. This is useful when considering the average ellipsoid intersection timing uncertainty of nearby targets, which is close to a few months. We showed that the distance to the target is the main factor in its timing uncertainty, therefore longer time scale observations might be needed for targets farther from Earth. Conversely, we might not need a full year's worth of data for targets closer to Earth to cover the entire window of interest. Thus, future work might include targets with smaller uncertainties that need less coverage (i.e. less than 11 sectors). To do this, an analytical treatment of these uncertainties and their propagation is ideal, and a relevant topic for future work, since it would allow for calculation of these precisions in advance. We also recommend enforcing the continuity of these sectors, to avoid gaps in light curves during their windows of interest. Future work might also involve consideration of more targets which only have data available in the TESS Full-Frame Images (FFIs).

Throughout this work, we assumed that the window of interest for a given target would only be determined by the uncertainty in their crossing time with the Ellipsoid. This implies that a signal sent in coordination with SN\,1987A would be broadcast immediately after the extraterrestrial civilization first observed the event. It instead might be appropriate to offer some additional margin after the signal's expected time-of-arrival, assuming that there may be some delay between event observation and signal transmission. The ideal duration of this window of time is unknown, and any choice is bound to be somewhat arbitrary, however, in the context of TESS, we recommend analyzing one or two sectors beyond the upper bound given by the ellipsoid crossing time uncertainty.

The SETI Ellipsoid framework provides no specific information related to the kind of synchronized technosignature we expect to see. We suggest searching for a signal that mimics the event's shape in the light curve, which would be detectable by TESS, while traditional SETI searches lean toward narrowband radio frequencies that are outside of TESS's detection capabilities. Further insights related to what a synchronized beacon might look like --- in the optical band or otherwise --- are still needed \citep{Sheikh2,Davenport_2022}. A potential avenue for future work could be refining a machine learning classifier such as the one used in Section \ref{ssec: light curves} with a training set containing specific light curve behaviors associated with the SETI Ellipsoid (e.g., mimicry of a supernova light curve). Using Full Frame Image data might also be prefered in this instance since these data haven't gone through long-term systematic trend removal.

The SETI Ellipsoid method, jointly with Gaia distances, provides a straightforward and flexible method for SETI searches that can be adapted to fit different modern surveys and source events. It can be applied retroactively to look for signals in archival data, as well as propagated forward in time to select targets and schedule monitoring campaigns as demonstrated by \cite{sn2023}. In this work, we have presented an approach that can be used with larger, upcoming time domain surveys such as LSST.

\begin{acknowledgements}

The authors thank Andy Nilipour for helpful discussions related to the SETI Ellipsoid method and its timing uncertainties. 

The authors acknowledge support from the Breakthrough Listen initiative. Breakthrough Listen is managed by the Breakthrough Initiatives, sponsored by the Breakthrough Prize Foundation.

JRAD acknowledges support from the DiRAC Institute in the Department of Astronomy at the University of Washington. The DIRAC Institute is supported through generous gifts from the Charles and Lisa Simonyi Fund for Arts and Sciences, and the Washington Research Foundation.

SC acknowledges support as the Director of the Berkeley SETI Research Center Research Experience for Undergraduates Site, supported by the National Science Foundation under Grant AST 1950897.

S.Z.S. acknowledges that this material is based upon work supported by the National Science Foundation MPS-Ascend Postdoctoral Research Fellowship under Grant No. 2138147.

We acknowledge ESA Gaia, DPAC and the Photometric Science Alerts Team (http://gsaweb.ast.cam.ac.uk/alerts).

Funding for the TESS mission is provided by NASA's Science Mission directorate.

This research made use of the cross-match service provided by CDS, Strasbourg. \citep{xmatch,xmatch2}

This research made use of Lightkurve, a Python package for Kepler and TESS data analysis \citep{lightkurve}.
\end{acknowledgements}

\software{
Python, IPython \citep{ipython}, 
NumPy \citep{numpy}, 
Matplotlib \citep{matplotlib}, 
SciPy \citep{scipy}, 
Pandas \citep{pandas}, 
Astropy \citep{astropy:2013,astropy:2018, astropy:2022}
Scikit-Learn \citep{scikit-learn}
}

\begin{table*}[ht!]
\begin{tabular}{|l|l|l|l|l|}
\hline
\textbf{TICID} & $\mathbf{RA^{(\circ)}}$ & $\mathbf{Dec^{(\circ)}}$ & $\mathbf{d_1(pc)}$ & $\mathbf{t_{cross}(yr)}$ \\ \hline
279055252 & 102.35 $\pm$ 0.02& -57.57 $\pm$ 0.02& 313.32$_{-1.81}^{+1.47}$ & 2018.58 $\pm$ 0.18\\
382187278 & 81.26 $\pm$ 0.01& -57.24 $\pm$ 0.01& 431.33$_{-2.13}^{+1.98}$ & 2018.60 $\pm$ 0.16\\
349374407 & 110.09 $\pm$ 0.01& -61.65 $\pm$ 0.01& 368.79$_{-1.72}^{+1.66}$ & 2018.78 $\pm$ 0.15\\
350434130 & 85.04 $\pm$ 0.01& -55.99 $\pm$ 0.01& 359.73$_{-1.69}^{+1.33}$ & 2018.78 $\pm$ 0.14\\
38711529 & 67.05 $\pm$ 0.01 & -60.36 $\pm$ 0.01 & 493.05$_{-1.93}^{+2.64}$ & 2018.89 $\pm$ 0.15 \\
350580827 & 86.89 $\pm$ 0.02& -55.04 $\pm$ 0.02& 313.21$_{-1.91}^{+2.22}$ & 2018.98 $\pm$ 0.23\\
306672432 & 120.59 $\pm$ 0.02& -67.79 $\pm$ 0.02& 360.68$_{-2.74}^{+2.62}$ & 2018.99 $\pm$ 0.25\\
260368525 & 95.24 $\pm$ 0.02 & -58.35 $\pm$ 0.02 & 445.65$_{-3.35}^{+4.58}$ & 2019.03 $\pm$ 0.32 \\
382067226 & 79.61 $\pm$ 0.01& -55.79 $\pm$ 0.01& 346.06$_{-1.18}^{+1.09}$ & 2019.08 $\pm$ 0.11\\
279614617 & 106.30 $\pm$ 0.01& -57.57 $\pm$ 0.01& 277.66$_{-0.88}^{+0.97}$ & 2019.14 $\pm$ 0.11\\
350711056 & 88.26 $\pm$ 0.01& -57.73 $\pm$ 0.01& 469.82$_{-3.03}^{+2.62}$ & 2019.24 $\pm$ 0.22\\
381976956 & 77.55 $\pm$ 0.01& -57.59 $\pm$ 0.01& 446.51$_{-2.93}^{+2.20}$ & 2019.28 $\pm$ 0.19\\ \hline
382028301 & 78.59 $\pm$ 0.01& -56.03 $\pm$ 0.01& 371.06$_{-1.47}^{+1.45}$ & 2020.64 $\pm$ 0.14\\
260416034 & 95.65 $\pm$ 0.04& -56.81 $\pm$ 0.04& 368.27$_{-4.47}^{+3.91}$ & 2020.65 $\pm$ 0.44\\
272316124 & 118.76 $\pm$ 0.01& -74.05 $\pm$ 0.01& 485.10$_{-3.10}^{+3.32}$ & 2020.69 $\pm$ 0.24\\
382028819 & 78.50 $\pm$ 0.02& -57.58 $\pm$ 0.02& 472.45$_{-3.24}^{+3.18}$ & 2020.75 $\pm$ 0.24\\
349681799 & 112.53 $\pm$ 0.02& -61.23 $\pm$ 0.02& 333.39$_{-1.77}^{+2.19}$ & 2020.83 $\pm$ 0.21\\
220404759 & 69.86 $\pm$ 0.01& -59.91 $\pm$ 0.01& 550.05$_{-3.26}^{+2.96}$ & 2020.87 $\pm$ 0.21\\
349571548 & 111.68 $\pm$ 0.01& -62.16 $\pm$ 0.01& 382.82$_{-1.58}^{+1.70}$ & 2020.89 $\pm$ 0.15\\
260266066 & 93.75 $\pm$ 0.01& -59.73 $\pm$ 0.01& 620.03$_{-5.94}^{+7.12}$ & 2020.95 $\pm$ 0.38\\
348900215 & 106.21 $\pm$ 0.02& -60.52 $\pm$ 0.02& 414.98$_{-3.23}^{+3.46}$ & 2020.96 $\pm$ 0.29\\
350715741 & 88.41 $\pm$ 0.02& -57.09 $\pm$ 0.02& 444.37$_{-2.65}^{+2.47}$ & 2021.01 $\pm$ 0.22\\
453100169 & 116.22 $\pm$ 0.02& -68.76 $\pm$ 0.02& 516.53$_{-4.82}^{+5.34}$ & 2021.10 $\pm$ 0.35\\
260542038 & 96.99 $\pm$ 0.01& -58.14 $\pm$ 0.01& 437.02$_{-2.25}^{+2.00}$ & 2021.17 $\pm$ 0.18\\
300869262 & 117.18 $\pm$ 0.01& -70.14 $\pm$ 0.01& 520.21$_{-2.97}^{+2.90}$ & 2021.27 $\pm$ 0.21\\
150431816 & 97.36 $\pm$ 0.01& -61.04 $\pm$ 0.01& 686.37$_{-4.98}^{+4.84}$ & 2021.29 $\pm$ 0.26\\
382159540 & 81.02 $\pm$ 0.01& -56.50 $\pm$ 0.01& 415.38$_{-1.74}^{+2.31}$ & 2021.32 $\pm$ 0.18\\
278731442 & 99.56 $\pm$ 0.01& -55.79 $\pm$ 0.01& 297.63$_{-1.11}^{+0.91}$ & 2021.37 $\pm$ 0.13\\
278684349 & 99.23 $\pm$ 0.01& -57.01 $\pm$ 0.01& 350.63$_{-1.33}^{+1.32}$ & 2021.40 $\pm$ 0.14\\
220424430 & 72.28 $\pm$ 0.01& -57.46 $\pm$ 0.01& 415.09$_{-2.21}^{+2.05}$ & 2021.42 $\pm$ 0.19\\
349972412 & 114.47 $\pm$ 0.01& -62.08 $\pm$ 0.01& 336.72$_{-1.25}^{+1.39}$ & 2021.43 $\pm$ 0.15\\
350954265 & 90.81 $\pm$ 0.01& -55.48 $\pm$ 0.01& 344.91$_{-1.00}^{+1.02}$ & 2021.50 $\pm$ 0.11\\ \hline
\end{tabular}
\caption{All 32 targets of interest found within 0.5\,ly of the SN\,1987A Ellipsoid at the mid-time of their TESS observing cycle, arranged chronologically by crossing time with the Ellipsoid. The twelve targets on the top section of the table were on the Ellipsoid in Cycle 1. The twenty targets on the bottom section of the table belong to Cycle 3. The TICID column shows the TESS ID for each object. $RA$ and $Dec$ values and uncertainties were obtained from TESS and Gaia data. The distances, \textbf{$d_1$}, and their uncertainties were retrieved from \cite{Bailer-Jones}.The Ellipsoid crossing times, $t_{cross}$, and their uncertainties were computed using the methods described in Section \ref{sec:select}. All values were rounded to the hundredth decimal place for visual simplicity and consistency.}
\label{bigtbl}
\end{table*}

\bibliography{bibliography.bib}
\bibliographystyle{aasjournal}

\end{document}